\documentclass{mystyle}

\usepackage[varg]{txfonts}
\usepackage{epsfig}

\renewcommand{\arraystretch}{0.8}

\def\slash#1{\setbox0=\hbox{$#1$}\dimen0=\wd0
      \setbox1=\hbox{/} \dimen1=\wd1 \ifdim\dimen0>\dimen1
      \rlap{\hbox to \dimen0{\hfil/\hfil}} #1                        \else
      \rlap{\hbox to \dimen1{\hfil$#1$\hfil}}
      /   \fi}
\def\simge{\mathrel{\rlap{\raise 0.511ex \hbox{$>$}}{\lower 0.511ex
\hbox{$\sim$}}}}
\def\simle{\mathrel{\rlap{\raise 0.511ex \hbox{$<$}}{\lower 0.511ex
\hbox{$\sim$}}}}
\def\slash#1{\setbox0=\hbox{$#1$}\dimen0=\wd0
      \setbox1=\hbox{/} \dimen1=\wd1 \ifdim\dimen0>\dimen1
      \rlap{\hbox to \dimen0{\hfil/\hfil}} #1                        \else
      \rlap{\hbox to \dimen1{\hfil$#1$\hfil}}
      /   \fi}

\newcommand{\be}{\begin{equation}}
\newcommand{\ee}{\end{equation}}
\newcommand{\bea}{\begin{eqnarray}}
\newcommand{\eea}{\end{eqnarray}}
\newcommand{\nn}{\nonumber}

\newcommand{\gev}{\,{\rm GeV}}
\newcommand{\kd}{K_{\ell 2}}
\newcommand{\kl}{K_{\ell 3}}
\newcommand{\vus}{\vert V_{us}\vert}
\newcommand{\vud}{\vert V_{ud}\vert}
\newcommand{\vub}{\vert V_{ub}\vert}
\newcommand{\fp}{f_+(0)}

\newcommand{\vusfpexp}{0.2160\pm 0.0005}
\newcommand{\vusLR}{0.2248\pm 0.0018_{\fp}\pm 0.0005_{exp}}
\newcommand{\vusfpLR}{0.2177\pm 0.0028}

\begin{document}

\title{$V_{us}$ from ${\kl}$ Decays}

\author{Federico~Mescia}
\address{Dip. di Fisica, Univ. degli Studi
          Roma Tre, Via della Vasca Navale 84, I-00146 Roma,
          Italy\\INFN, Laboratori Nazionali di Frascati, Via E.
          Fermi 40, I-00044 Frascati, Italy}
\twocolumn[\maketitle\abstract{
Important progress made this year,  both in theory and in experiment, helped solving the 
problem of $2\sigma$--deviation from the unitarity of the first row elements in the CKM matrix. 
Today we have,  $\vert V_{us}\vert^2+\vert V_{ud}\vert^2 +\vert
V_{ub}\vert^2-1=-0.0008(13)$, $-0.0010(13)$, or $-0.0005(13)$, depending on whether the $q^2$-dependence
of the relevant $\kl$ form factor is considered as pole-like, linear or quadratic function, and on
the Leutwyler-Roos value of $f_+(0)=0.961(8)$, whose validity
has recently been reinforced by lattice studies.
In this talk  we summarize the
recent  developments. }]

\maketitle

\section{Introduction}

$\vud$ and $\vus$ are fundamental parameters of the Standard Model. 
The Cabibbo--Kobayashi-Maskawa (CKM) unitarity implies that  
 $\vert V_{us}\vert^2+\vert V_{ud}\vert^2 +\vert V_{ub}\vert^2=1$. 
 In this equality, $\vub$ is negligible in size, whereas  $\vud$ and $\vus$ induce 
 comparable uncertainties.

The two most important determinations of $\vud$ come from nuclear 
$0^+\to 0^+$ transitions, and from the neutron beta decays.
With respect to the value quoted in PDG~\cite{PDG}, a sign error  
of the radiative corrections to the neutron beta decays has been recently 
corrected~\cite{marciano}. Consequently, the updated average value for $\vert V_{ud}\vert$ 
now reads:
\be
\label{eq:vud}
|V_{ud}| = 0.9740 \pm 0.0005.
\ee
Using this value and by imposing the CKM unitarity,  the Cabibbo angle ($\vert V_{us}\vert$)  
amounts to
\be
\label{eq:vusuni}
|V_{us}|^{\rm{uni.}} = 0.2265 \pm 0.0022.
\ee
Testing the unitarity of the $1^{st}$-row of the CKM matrix means a comparison of this value 
with $\vus$ deduced directly from the processes governed by the $s\to u$ transition.
Although theoretical constraints on $\vus$ from  the semileptonic hyperon decays~\cite{Cabibbo:2003cu,Flores-Mendieta:2004sk,Guadagnoli:2004qw}, 
$\tau \to K \nu_\tau$~\cite{Gamiz:2004ar} and leptonic kaon decays~\cite{milc,marciano} ($\kd$)
recently became promising too, the best determination of $\vus$
is still obtained from $K\to\pi\ell\nu$ decay modes ($\kl$).

Before concentrating on the semileptonic $\kl$ decay,  it is important to mention the intensive 
activity within the lattice QCD community invested in reducing the errors on the estimate of
$f_K/f_\pi$ (cfr ref.~\cite{milc}).  Once combined with the experimentally established $\Gamma(K\to
\mu \nu)/\Gamma(\pi\to \mu \nu)$, this would allow for a  precise determination of $\vert
V_{us}/V_{ud}\vert$, and thus of $\vert V_{us}\vert$.  This is why the experimenters  recently 
became more interested in increasing the accuracy in measuring the  $K_{\mu 2}$ decay
rates~\cite{KLOE}. It should be stressed, however,  that the current accuracy  on $f_K/f_\pi-1$ is
about $6.5\%$, which amounts to a relative error of $1.2\%$ for $\vus$.  Therefore to achieve the challenging 
 $\delta\vus/\vus=0.1\%$, the relative error of  $f_K/f_\pi-1$ should be  $0.5\%$ or less,
which is hardly feasible.  In $\kl$ decay, instead,
the equivalent requires a theoretical  uncertainty of $7\%$, thanks to the conservation  of the vector
current (CVC) and the Ademollo--Gatto theorem (AGT)~\cite{ag}. Such an accuracy is within  reach for
the forthcoming lattice QCD studies. 

\section{{$\kl$} decay modes and $\vus$}
\label{sec:formula}

\begin{figure}[bht]
\hspace*{-0.95cm}
\includegraphics[width=0.58\textwidth,height=0.5\textwidth]{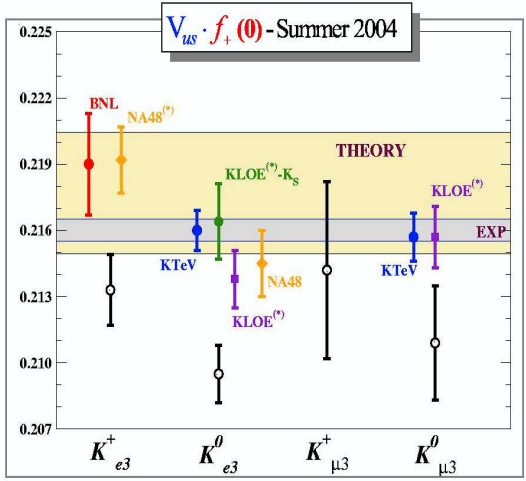}
\caption{ Results of $\vus  \fp$. Full points are obtained by assuming 
the pole-fit choice and recent measures (see table\ref{tab:final}).
Empty points are  on the older experiments~\protect\cite{PDG}. 
The ``EXP" and ``THEORY" bands indicate respectively the average of the new 
experimental results (eq.~(\ref{eq:expquad})) 
and the unitarity prediction in eq.~(\ref{eq:uniband}).} 
\label{fig:fvus}
\end{figure}

We first recall to the master formula for the $\kl$ decay rate: 
\bea
\Gamma(K_{\ell 3(\gamma)}) &=& 
{ G_F^2 M_K^5 \over 128 \pi^3} C_K^2   
  S_{\rm ew}\,|V_{us}|^2 f_+(0)^2\\
&&I_K^\ell(\lambda_{+,0})\,\left(1 + \delta^{K}_{SU(2)}+\delta^{K \ell}_{\rm
em}\right)^2\,.
\nn\label{eq:one}
\eea
$C_{K}^2$ is  equal  to 1 ($1/2$) for the neutral (charged) kaon decay; 
$I_K^\ell(\lambda_{+,0})$ is the phase space integral defined in absence 
of electromagnetic corrections and depending on the slope parameters $\lambda_{+,0}$ 
which will be discussed below;  
$S_{\rm ew}=1.0232(3)$ is the universal short-distance electromagnetic 
correction~\cite{Sirlin:1981ie}  evaluated at  $\mu=M_\rho$;  
$\delta^{K \ell}_{\rm em}$ and  $\delta^{K}_{SU(2)}$ are respectively  the 
long-distance electromagnetic and strong isospin-breaking corrections; finally,  $f_{+}(0)$ is 
the vector form factor at  zero momentum transfer [$t\equiv q^2=(p_K-p_\pi)^2 = 0$] 
which encodes the SU(3)  breaking effects in the hadronic matrix element.
\begin{table*}[htb]
\setlength{\tabcolsep}{3.5pt}
\renewcommand{\arraystretch}{1.0} 
\centering
\footnotesize
\begin{tabular}{c||c|c||c|c|c|c||c|c|}
&\multicolumn{2}{|c||}{$K^+_{e3}$} 
& \multicolumn{3}{|c|}{$K^L_{e3}$}& $K^S_{e3}$
& \multicolumn{2}{|c|}{$K^L_{\mu3}$} \\
\cline{2-9}
& BNL & NA48$^{(*)}$ & KTeV  & KLOE$^{(*)}$ & NA48 & KLOE$^{(*)}$ &KTeV & KLOE$^{(*)}$  \\ 
\hline
Br[\%]& $5.13(10)$ & $5.14(6)$ & $40.67(11)$ & $39.85(35)$ & $40.10(45)$ & $0.0709(11)$ & $27.01(9)$ & $27.02(25)$ \\ 
\hline
& \multicolumn{8}{|c|}{Linear Parameterization}\\
\hline
$I^\ell_K$ & \multicolumn{2}{|c||}{$0.10627(15)$} 
& \multicolumn{4}{|c||}{$0.10337(15)$}
& \multicolumn{2}{|c|}{$0.06877(16)$}\\
\hline
$\vus \fp$& $0.2185(23)$ & $0.2187(15)$ & $0.2155(9)$&$0.2133(13)$ &$0.2140(15)$ &$0.2160(16)$
&$0.2148(10)$ &$0.2148(14)$ \\ 
\hline
& \multicolumn{8}{|c|}{Pole Parameterization}\\
\hline
$I^\ell_K$ & \multicolumn{2}{|c||}{$0.10580(15)$} 
& \multicolumn{4}{|c||}{$0.10291(15)$}
& \multicolumn{2}{|c|}{$0.06820(18)$}\\
\hline
$\vus \fp$& $0.2190(23)$ & $0.2192(15)$ & $0.2160(9)$&$0.2138(13)$ &$0.2145(15)$ & $0.2164(17)$ & $0.2157(11)$ &$0.2157(14)$ \\ 
\hline
&\multicolumn{8}{|c|}{Quadratic Parameterization}\\
\hline
$I^\ell_K$ & \multicolumn{2}{|c||}{$0.10520(71)$} 
& \multicolumn{4}{|c||}{$0.10233(70)$}
& \multicolumn{2}{|c|}{$0.06777(48)$}\\
\hline
$\vus \fp$&$0.2196(24)$ & $0.2198(17)$ & $0.2166(12)$ & $0.2144(15)$&$0.2151(17)$ &
$0.2171(17)$ & $0.2164(13)$ & $0.2164(16)$ \\ 
\hline
\end{tabular}
\caption{  Recent 
 results from BNL-E865~\protect\cite{BNL}, 
 KTeV~\protect\cite{KTeV}, NA48~\protect\cite{NA48} and KLOE~\protect\cite{KLOE} and corresponding values of $\vus \fp$. 
Preliminary results are marked by $(*)$ . We use for the linear parametrization
 $\lambda_{+}=0.0281(4)$ and $\lambda_{0}=0.017(1)$, for the pole one 
 [$f_{+,0}(t)= f_{+}(0)/( 1 - \lambda_{+,0}\, t/m^2_{\pi^+})$], $\lambda_{+}=0.0250(4)(4)$ and 
$\lambda_{0}=0.014(1)$ 
 from~\protect\cite{KTeV}, and for the quadratic one [$f_{0}(t)= f_{+}(0)( 1 + \lambda_{0}\, t/m^2_{\pi^+})$, and
$f_{+}(t)= f_{+}(0) ( 1 + \lambda'_{+}\, t/m^2_{\pi^+} +\lambda''_{+}\,t^2/(2\,m^4_{\pi^+})$]    
$\lambda_{0}=0.0137(13)$, $\lambda'_{+}=0.0206(18)$, and $\lambda''_{+}=0.0032(7)$ from~\protect\cite{KTeV}.
In addition, 
$\tau^{PDG}_{K_L}=5.15(4)\times 10^{-8}\,s$,  
$\tau^{PDG}_{K^+}=1.2384(24)\times 10^{-8}\,s$ and 
$\tau^{PDG}_{K_S}=8.953(8)\times 10^{-11}\,s$  are used along with 
 $\delta^{K \ell}_{\rm em}$   for the fully inclusive rate. }
\label{tab:final}
\end{table*}
\renewcommand{\arraystretch}{0.9} 
To extract the value of $\vus$ from eq.~(\ref{eq:one}) one needs  not only an accurate 
experimental values for the rate ($\Gamma$) and for the kinematic integral $I_K^\ell$,  but also 
the theoretical estimates of the $\delta$'s and $\fp$. In what follows,  we provide the update 
to each of these quantities.\\
{\bf Width measurements:}
This summer, all the new generation kaon
experiments released  results for the $\kl$ decay modes. The important novelty is that these new
results  are consistent among themselves (see  table~\ref{tab:final}),  but they  
disagree with the old ones. \\
{\bf $\mathbf{I_K^\ell(\lambda_{+,0})}$ and the form factor shapes:}
KTeV~\cite{KTeV}, ISTRA+~\cite{ISTRA+} and NA48~\cite{NA48} studied the $t$-de\-pen\-den\-ce of the 
partial $\kl$ rates: KTeV and ISTRA+ analyzed both muonic and electronic  decays, while  
NA48 restrained to $K^L_{e3}$ only. Dalitz plot data have 
been examined by assuming a linear, quadratic and pole dependence in $t$.
With the linear function, $f_{+,0}(t)=f_{+}(0)( 1 + \lambda_{+,0}\, t/m^2_{\pi^+})$, 
the three groups agree on the values for the slopes, $\lambda_{+,0}$, which are more accurate 
than the ones reported by the PDG. 
The average is  $\lambda_{+}=0.0281(4)$, and $\lambda_{0}=0.017(1)$. 
Concerning the presence of the quadratic term in $f_{+}(t)$, 
findings are controversial: contrary to  NA48, KTeV and   ISTRA+ collaborations observe 
a non-zero curvature, but with $1 \sigma$ significance only. A closer look at 
the systematics, and the results by  KLOE are certainly needed. Finally,  the pole fit, tried by both 
KTeV and NA48, looks, for the time being, the most reasonable solution  and the measured pole mass
is consistent  with the mass of $K^*(892)$  (as anticipated by the lattice study, earlier this year~\cite{noi}). 
In the case of the scalar form factor $f_{0}(t)$
no curvature has been observed.\\
{\bf Strong and em isospin breaking effects:}
%
\begin{table}[htb]
\setlength{\tabcolsep}{3.8pt}
\centering
\begin{tabular}{c||c||c|c}
& $\delta^K_{SU(2)} (\%)$ 
& \multicolumn{2}{|c}{
$\delta^{K \ell}_{\rm em}(\%) $}    \\
&  &
\multicolumn{2}{|c}{ 3-body $\qquad\qquad$ full}  \\
\hline
$K^{+}_{e3}$    & 2.31(22)  &  -0.35(16)$\phantom{[*]}$ & -0.10(16)$\phantom{[*]}$ \\
$K^{0}_{e 3}$   &  0        &  +0.30(10)$\phantom{[*]}$ & +0.55(10)$\phantom{[*]}$ \\
$K^{+}_{\mu 3}$ & 2.31(22)  &  -0.05(20)[*] & +0.20(20)[*]\\
$K^{0}_{\mu 3}$ &  0        &  +0.55(20)$\phantom{[*]}$ & +0.80(20)$\phantom{[*]}$\\
\end{tabular}
\caption{ Summary of the isospin-breaking
factors~\protect\cite{Cirigliano,Andre,ginsberg}: 
$\delta^{K \ell}_{\rm em}$ [3 body] denotes corrections for
 the inclusive rate 
involving  radiative events inside 
the $K_{\ell 3}$ Dalitz Plot, whereas
$\delta^{K \ell}_{\rm em}$ [full]  those 
for the fully inclusive $K_{\ell 3(\gamma)}$ rate. 
The entries with $[*]$ are from ref.~\protect\cite{ginsberg}.}
\label{tab:iso-brk}
\end{table}
$\delta^{K \ell}_{\rm em}$ and $\delta^{K}_{SU(2)}$ corrections 
 have been recently and properly calculated  at ${\cal O}[(m_d-m_u)p^2,e^2 p^2]$ in~\cite{Cirigliano,Andre}. 
The numerical results are collected  in table~\ref{tab:iso-brk}.\\
{\bf $\mathbf{\vus \fp}$ estimates:}
With the three in\-gre\-di\-ents discussed so far, we can extract   $\vus \fp$  
with small theoretical errors  from  both   charged and
neutral modes (see table~\ref{tab:final}), allowing us a  first consistency  check~\cite{Cirigliano} between experiment
and theory. 
In fig.~\ref{fig:fvus} we show the points obtained by assuming the 
pole-like $t$-dependence for the form factors with the corresponding pole masses 
determined by KTeV.  
The resulting experimental average reads 
\be
(\vus \fp)^{exp}=
\vusfpexp\,,
\label{eq:expquad}
\ee
which is represented  in fig.~\ref{fig:fvus} by the dark-shaded band. 
Had we used the linear (quadratic) pa\-ra\-me\-te\-ri\-za\-tion, the central value would shift  
by $-0.07\%$ ($+0.06\%$). 
At this conference it was argued~\cite{KLOE} that the small difference between the values 
of $\vus \fp$  as extracted from  $\kl^+$ and  from $\kl^0$ might be 
due to a problem of the present value of the $K_L$-lifetime.\\ 
{\bf SU(3) breaking effects and $\mathbf{\fp}$:}
The remaining ingredient to extract $\vus$ from eq.~(\ref{eq:expquad}) is $\fp$. 
This quantity is the origin of the largest   uncertainty in $\vus$, namely  
$\delta\vus/\vus\simeq 1\%$, to be compared with $0.2\%$ and  $0.35\%$ 
coming from the isospin breaking corrections and the uncertainty on the phase space integral, 
respectively. According to eq.(\ref{eq:one}), $\fp$  is defined in the absence
of em and strong isospin-breaking terms and incorporates
only strong SU(3)-breaking effects. 
Its expansion in chiral perturbation theory (ChPT) reads,
 \be\label{chpt4}
 f_+(0)= 1 + f_2 + f_4 + \ldots,
 \ee
where $f_+(0)= 1$ reflects the CVC in the SU(3) limit, while $f_2$ and $f_4$ stand for the leading and 
next-to-leading chiral corrections.  AGT ensures that the SU(3) breaking corrections are quadratic in 
$(m_s-m_u)$ and $f_2=-0.023$ is a clean prediction by ChPT, i.e.  
no unknown couplings enter at ${\cal O}(p^4)$. 
The calculation of the chiral loop contribution, $\Delta(\mu)$ in
\be 
\renewcommand{\arraystretch}{0.5} 
f_4 =
\Delta(\mu) + f_4\vert^{loc}(\mu)\,, 
\label{eq:f4ch}
\ee 
has been recently completed~\cite{tal,post}. The estimate of $f_4$, however, 
still suffers from the  uncertainty due to the lack of knowledge of the low energy constants 
entering  $f_4\vert^{loc}(\mu)$. The
PDG quotes the value obtained in the quark-model  
cal\-cu\-la\-tion by Leutwyler-Roos~\cite{LR}~(LR),  
\be 
f_4=-0.016\pm 0.008 \to f_{+}(0)=0.961(8), \label{eq:LR} 
\ee 
based on parameterization of the asymmetry between kaon and pion
wave functions.  If the esti\-ma\-te of $\fp$ in eq.~(\ref{eq:LR}) is used along with the experimental 
ave\-ra\-ge of $\vus \fp$, eq.~(\ref{eq:expquad}),  one gets 
\be
\vus=\vusLR\,,
\ee
in good agreement with the value obtained by imposing the CKM unitarity [cfr eq.~(\ref{eq:vusuni})].  
This com\-pa\-ti\-bi\-li\-ty is also  observed in fig.~\ref{fig:fvus} where the light-shaded band   refers 
to, 
\be
\vert V_{us}^{uni.}\vert  \fp^{theo.}_{eq.(\ref{eq:LR})}=\vusfpLR\,.
\label{eq:uniband}
\ee
The LR estimate has been corroborated this year  by a (quenched) lattice QCD study that gave~\cite{noi}
\be
f_4=-0.017\pm 0.009 \to f_{+}(0)=0.960(9).
\ee
In this estimate, the leading quenched artifacts have been subtracted, but residual effects at  ${\cal
O}(p^6)$ are still present. A conservative uncertainty of $ 60\%$ 
 has been attributed to  $f_4$~\cite{noi}, which can be substantially reduced by
 an unquenched  
 calculation. Besides the lattice estimates, two more calculations~\cite{tal,jam} appeared this year, yielding respectively:
\bea 
\renewcommand{\arraystretch}{0.5} 
\!\!\!\!\!\!\!\!\!f_4&=&-0.001\pm 0.010 \to
f_{+}(0)=0.976(10)\,,\label{eq:tal}\\ 
\!\!\!\!\!\!\!\!\!f_4&=&-0.003\pm 0.011 \to
f_{+}(0)=0.974(11)\,.
\label{eq:jam} 
\eea 
However they both contain model-dependent assumptions. 
In particular  a strong ansatz to get $f_4\vert^{loc}$  
has been imposed in~\cite{tal}, which then propagates to ref.~\cite{jam} where the value~(\ref{eq:tal}) is used as in\-put.  
Specifically, the authors of~\cite{tal}  identify the LR value of $f_4=-0.016(8)$ with $f_4\vert^{loc}(\mu)$ 
at the scale $\mu=m_{\rho}$. For the estimate  of $\fp$ this means  
adding
 the loop contribution $\Delta(\mu=m_{\rho})=0.015$  to the LR value.
  Such an interpretation of the LR result is questionable:  
the choice  $\fp=0.961(8)+\Delta(\mu)$ could be carried
out at a different scale.  In this case, by varying $\mu$ in a reasonable range 
[$0.5\gev$-$1\gev$]: $\Delta(\mu)=3.5\%\to 0.4\%$ and $\fp=0.996(8)\to 0.965(8)$. Because of this scale
uncertainty,  the error bars in eqs.~(\ref{eq:tal},\ref{eq:jam}) should be considerably larger  (see comment
in~\cite{Cirigliano}).   
Notice also that by using the values of $\fp$~(\ref{eq:tal}) and (\ref{eq:jam}), 
unitarity is violated by about $+1.4\,\sigma$. The corresponding $\vus \fp$
theory band in fig.~\ref{fig:fvus} would be shifted to $\vert V_{us}^{uni}\vert\fp =0.221(3)$, i.e. consistent with the $K^+$
experimental values, but well above the $K^0$ ones. 

Before concluding we should stress that in literature~\cite{NA48,ckmfitter}   
 the value  
$f^{K^0 \pi^-}(0)=0.981\pm 0.010$, and $f^{K^+ \pi^0}(0)=1.002\pm 0.010$ 
of ref.~\cite{Cirigliano}  are  erroneously treated as  independent estimates of  $\fp$,  
and directly compared  to the ones discussed in this write-up. The apparent inconsistency is due to 
the fact that the above results refer to a different definition of $\fp$, in which some of the isospin breaking 
corrections are included in the definition.  Once we remove these corrections to perform 
 a consistent comparison with the standard definition  
(used in this write-up, by the PDG~\cite{PDG} and by KTeV~\cite{KTeV}), the two above values give 
$\fp=0.976(10)$, which is the result quoted in eq.~(\ref{eq:tal}). 
  Our analysis of  $\vus$ gives exactly 
the same result of~\cite{Cirigliano}  as long as the
  SU(3)-breaking estimates  in the form factors  are kept identical.

In conclusion,  a novel route to estimate $f_4$  by means of lattice QCD has been devised this year.  
The quenched value essentially confirms the one obtained long ago by LR. This result and, more importantly, the new 
experimental data helped resolving the puzzle  of  the $1^{st}$ row  CKM-unitarity.  
To perform a more accurate test of  CKM unitarity, an unquenched 
lattice QCD calculation of $f_4$ is needed.  In a less near future, an alternative will 
be the proposal of ref.~\cite{tal}, who showed 
that the couplings in $f_4$ can be determined from the precision measurement of  the slope and curvature 
of the scalar form factor $f_0(t)$. 

\section*{ACKNOWLEDGMENTS}
We thank   M.~Antonelli, P.~Franzini, G~Isidori and  
A.~Sibidanov for discussions on the subject  of this talk. 
I am also very grateful to all my friends and colleagues  of the 
SPQ$_\mathrm{CD}$R  Collaboration.
The work of  F.M.~is partially supported by 
IHP-RTN, EC contract No.~HPRN-CT-2002-00311 (EURIDICE).


\end{document}